\documentclass[manuscript,natbib,screen=true]{acmart}

\PassOptionsToPackage{dvipsnames}{xcolor}
\usepackage{dsfont}
\usepackage{acronym}
\usepackage{amsmath}
\usepackage[linesnumbered,lined,ruled,commentsnumbered]{algorithm2e}
\usepackage{bbm}
\usepackage{bm}
\usepackage{booktabs}
\usepackage[skip=0pt]{caption}
\usepackage[T1]{fontenc}
\usepackage{graphicx}
\usepackage{hyperref}
\usepackage{listings}
\usepackage{multirow}
\usepackage{natbib}
\usepackage{subcaption}
\usepackage{tabularx}
\usepackage[dvipsnames]{xcolor}
\usepackage{enumerate}
\usepackage[inline]{enumitem}
\usepackage{mleftright}
\usepackage{xcolor, soul}
\usepackage{tablefootnote}

\acrodef{GNN}{graph neural network}
\acrodef{IR}{information retrieval}
\acrodef{LTR}{learning to rank}
\acrodef{MAB}{multi-armed bandit}
\acrodef{MC}{Markov chain}
\acrodef{NBR}{next basket recommendation}
\acrodef{NBRR}{next basket repurchase recommendation}
\acrodef{NNBR}{next novel basket recommendation}

\acrodef{PIF}{personal frequency information}
\acrodef{RNN}{recurrent neural network}
\acrodef{TREx}{two-step repetition-exploration}
\acrodef{TREx-Rep}{TREx with repetition module only}
\acrodef{BERT4Expl}{BERT for exploration in grocery shopping}
\acrodef{SOTA}{state-of-the-art}

\acrodef{BTBR}{bi-directional transformer basket recommendation}

\newcommand{\header}[1]{\vspace{1mm}\noindent\textbf{#1.}}

\definecolor{paleyellow}{HTML}{FFEF77}
\definecolor{paleorange}{HTML}{FBB068}
\definecolor{paleblue}{HTML}{65B2FF}

\newcommand{\Better}[1]{\rlap{*}}

\author{Ming Li}
\orcid{0000-0001-7430-4961}
\affiliation{%
        \institution{University of Amsterdam}
        \city{Amsterdam}
        \country{The Netherlands}
}
\email{m.li@uva.nl}

\author{Mozhdeh Ariannezhad}
\orcid{0000-0002-1113-8094}
\affiliation{%
  \institution{AIRLab, University of Amsterdam}
  \streetaddress{Science Park}
  \city{Amsterdam}
  \country{The Netherlands}}
\email{m.ariannezhad@uva.nl}

\author{Andrew Yates}
\orcid{0000-0002-5970-880X} 
\affiliation{%
        \institution{University of Amsterdam}
        \city{Amsterdam}
        \country{The Netherlands}
}
\email{a.c.yates@uva.nl}

\author{Maarten de Rijke}
\orcid{0000-0002-1086-0202} 
\affiliation{%
        \institution{University of Amsterdam}
        \city{Amsterdam}
        \country{The Netherlands}
}
\email{m.derijke@uva.nl}

\copyrightyear{2023} 
\acmYear{2023} 
\setcopyright{rightsretained} 
\acmConference[RecSys '23]{Seventeenth ACM Conference on Recommender Systems}{September 18--22, 2023}{Singapore, Singapore}
\acmBooktitle{Seventeenth ACM Conference on Recommender Systems (RecSys '23), September 18--22, 2023, Singapore, Singapore}\acmDOI{10.1145/3604915.3608803}
\acmISBN{979-8-4007-0241-9/23/09}

\begin{document}

\title[Next Novel Basket Recommendation]{Masked and Swapped Sequence Modeling for Next Novel Basket Recommendation in Grocery Shopping}

\begin{abstract}

\Ac{NBR} is the task of predicting the next set of items based on a sequence of already purchased baskets.
It is a recommendation task that has been widely studied, especially in the context of grocery shopping.
In \ac{NBR}, it is useful to distinguish between repeat items, i.e., items that a user has consumed before, and explore items, i.e., items that a user has not consumed before. Most \acs{NBR} work either ignores this distinction or focuses on repeat items.
We formulate the \emph{next novel basket recommendation} (NNBR) task, i.e., the task of recommending a basket that only consists of \emph{novel items},
which is valuable for both real-world application and \acs{NBR} evaluation. 
We evaluate how existing \ac{NBR} methods perform on the \acs{NNBR} task and find that, so far, limited progress has been made w.r.t. the \acs{NNBR} task.
To address the \acs{NNBR} task, we propose a simple \textbf{b}i-directional \textbf{t}ransformer \textbf{b}asket \textbf{r}ecommendation model (BTBR), which is focused on directly modeling item-to-item correlations within and across baskets instead of learning complex basket representations.
To properly train BTBR, we propose and investigate several masking strategies and training objectives: 
\begin{enumerate*}[label=(\roman*)]
    \item item-level random masking,
    \item item-level select masking,
    \item basket-level all masking,
    \item basket-level explore masking, and
    \item joint masking. 
  \end{enumerate*}
In addition, an item-basket swapping strategy is proposed to enrich the item interactions within the same baskets.
We conduct extensive experiments on three open datasets with various characteristics. 
The results demonstrate the effectiveness of BTBR and our masking and swapping strategies for the \acs{NNBR} task. 
BTBR with a properly selected masking and swapping strategy can substantially improve \acs{NNBR} performance.
\end{abstract}



\maketitle

\acresetall


\section{Introduction}
Next basket recommendation is a type of sequential recommendation that aims to recommend the next basket, i.e., set of items, to users given their historical basket sequences. 
Recommendation in a grocery shopping scenario is one of the main use cases of the \acs{NBR} task, where users usually purchase a set of items instead of a single item to satisfy their diverse needs. 
Many methods, based on a broad range of underlying techniques (i.e., RNNs~\citep{dream, beacon, clea, sets2sets, bai2018attribute}, self-attention~\citep{dnntsp,sun2020timestamp2,chen2021modelingcat}, and denoising via contrastive learning~\citep{clea}), have been proposed for, and achieve good performance on, the \acs{NBR} task. 

\header{Next novel basket recommendation}
A recent study~\citep{nbr-rep-expl} offers a new evaluation perspective on the \ac{NBR} task by distinguishing between \emph{repetition} (i.e., recommending items that users have purchased before) and \emph{exploration} (i.e., recommending items that are new to the user) tasks in \acs{NBR} and points out the imbalance in difficulty  between the two tasks.
According to the analysis of existing methods in~\citep{nbr-rep-expl}, the performance of many existing NBR methods mainly comes from being biased towards (i.e., giving more resources to) the repetition task and sacrificing the ability of exploration. 
Building on these insights, recent work on \ac{NBR} has seen a specific focus on the pure repetition task \citep[e.g.,][]{buycycle} as well the introduction of specific methods for the repetition task~\citep{recanet, buycycle}.

\begin{table}
    \newcommand{\tabincell}[2]{\begin{tabular}{@{}#1@{}}#2\end{tabular}}
      \caption{Three types of basket recommendation.}
      \label{task-difference}
      \setlength{\tabcolsep}{2pt}
      \begin{tabular}{l llll}
        \toprule
        \bf Task & Target items  & Recommended basket & Related work\\
        \midrule
        NBR & Repeat items \& novel items & Repeat items \& novel items& \citep{dream, beacon, clea, sets2sets, bai2018attribute, dnntsp,sun2020timestamp2,chen2021modelingcat, clea}\\
        NBRR & Repeat items & Only repeat items & \citep{buycycle,recanet}\\
        NNBR & Novel items & Only novel items & This paper\\
        \bottomrule
      \end{tabular}
\end{table}
    
Novelty and serendipity are two important objectives when evaluating recommendation performance~\citep{kaminskas2016diversity, herlocker2004evaluating}. 
People might simply get tired of repurchasing the same set of items.
Even when they engage in a considerable amount of repetition, there is still a large proportion of users who would like to try something new when shopping for groceries~\citep{nbr-rep-expl}.
This phenomenon is especially noticeable for users with fewer transactions in their purchase history~\citep{recanet}.
Therefore, one of the key roles of recommender systems is to assist users in discovering potential novel items that align with their interests. However, in contrast to the pure repetition task, the pure exploration task in \acs{NBR} remains under-explored. 
Besides, due to the difference in difficulty between the two tasks, many online e-commerce and grocery shopping platforms have started to design a ``buy it again'' service to isolate repeat items from the general recommendation.\footnote{After login, users may see a ``buy it again'' page on e-commerce platforms (see, e.g., Amazon \url{https://amazon.com} and grocery shopping platforms (see, e.g., Picnic \url{https://picnic.app}), where the platform collects repeat items. 
Similarly, in the grocery shopping scenario, ``Try Something New'' services also exist, where only novel items are recommended to the user.}$^,$\footnote{See, e.g., \href{http://community.apg.org.uk/fileUploads/2007/Sainsburys.pdf}{http://community.apg.org.uk/fileUploads/2007/Sainsburys.pdf} for an example of the ``Try Something New'' concept in offline retail, and the Weekly New Recipe service at \href{https://ah.nl/allerhande/wat-eten-we-vandaag/weekmenu}{https://ah.nl/allerhande/wat-eten-we-vandaag/weekmenu} for an example in online retail.}

Motivated by the research gaps and real-world demands, we formulate the \acfi{NNBR} task, which focuses on recommending a novel basket, i.e., a set of items that are new to the user, given the user's historical basket sequence. 
Different from the repetition task, which predicts the probability of repurchase from a relatively small set of items, the \acs{NNBR} task needs to predict possible items from many thousands of candidates by modeling item-item correlations, which is more complex and difficult~\citep{nbr-rep-expl}.
\acs{NNBR} is especially relevant to the ``Try Something New''  concept in the grocery shopping scenario.
Table~\ref{task-difference} shows differences between three types of basket recommendation and positions our work.

\header{From NBR to NNBR} 
The \acs{NNBR} task can be seen as a sub-task of the conventional \acs{NBR} task, in which \acs{NBR} methods are designed to find all possible items (both \emph{repeat items} and \emph{novel items}) in the next basket. Therefore, it is possible to generate a novel basket by selecting only the top-$k$ novel items predicted by \acs{NBR} methods.
To modify \acs{NBR} methods for the \acs{NNBR} task, an intuitive solution is to remove the repeat items from the ground-truth labels and train models only depending on the novel items in the ground-truth labels.
Given this obvious strategy and given that many methods have already been proposed for \acs{NBR}, an important question is:
\emph{If we already have an \acs{NBR} model, do we need to train another model specifically for the \acs{NNBR} task?}
Surprisingly, we find that training specifically for exploration does not always lead to better performance in the \acs{NNBR} task, and might even reduce performance in some cases.

\header{BTBR: A bi-directional transformer basket recommendation method}
In \acs{NNBR}, item-to-item correlations are especially important, since we need to infer the utility of new items based on previously purchased items. 
Besides, a single basket is likely to address diverse needs of a user~\citep{triple2vec}. E.g., what a user would like to drink is more likely to depend on what he or she drank before rather than on the tooth paste they previously purchased.
However, most existing \acs{NBR} approaches~\citep{dream, beacon, sets2sets, dnntsp, clea} are two-stage methods, which first generate a basket-level representation~\citep{basket-emb}, and then learn a temporal model based on basket-level representations, which will lead to information loss w.r.t.\ item-to-item correlations~\citep{dnntsp, beacon,sun2020timestamp2}.
Some methods~\citep{dnntsp, beacon,sun2020timestamp2} learn partial item-to-item correlations based on the co-occurrence within the same or adjacent basket as auxiliary information beyond basket-level correlation learning. 
Instead of learning or exploiting complex basket representations, we learn item-to-item correlations from direct interactions among different items across different baskets. 
To do so, we propose a bi-directional transformer basket recommendation model (BTBR) that adopts a bi-directional transformer~\citep{vaswani2017attention} and uses the shared basket position embedding to indicate items' temporal information.

\header{Masking and training}
To properly train BTBR, we propose and investigate several masking strategies and training objectives at different levels and tasks, as follows: 
\begin{enumerate*}[label=(\roman*)]
\item item-level random masking: a cloze-task loss~\citep{devlin2018bert,taylor1953cloze}, in which we randomly mask the historical sequence at the item level;
\item item-level select masking: a cloze-task loss designed for exploration, in which we first select the items we need to mask and then mask all the occurrences of the selected item;
\item basket-level all masking: a general basket recommendation task loss, in which we mask and predict the complete last basket at the end of the historical sequence;
\item basket-level explore masking: an explore-specific basket recommendation task loss, in which we remove the repeat items and only mask the novel items in the last basket of the historical sequence; and
\item joint masking: a loss that follows the pre-train-then-fine-tune paradigm, in which we first adopt item-level masking for the cloze task, then fine-tune the model using basket-level masking.
\end{enumerate*}

In addition, conventional sequential item recommendation usually assumes that the items in a sequence are strictly ordered and sequentially dependent. 
However, recent work~\citep[e.g.,][]{flexible-order, cheng2021learning, xie2022contrastive, sequential-survey} argues that the items may occur in any order, i.e., the order is flexible, and ignoring flexible orders might lead to information loss.
Similarly, it is unclear whether the items that are being purchased across baskets have a strict order in the grocery shopping scenario. 
Thus, we propose an item swapping strategy that allows us to randomly move an item to another basket according to a certain ratio, which can enrich item interactions within the same basket.

We conduct extensive experiments on three publicly available grocery datasets to understand the effectiveness of the BTBR model and the proposed strategies on datasets with various repeat ratios and characteristics.

\header{Our contributions}
The main contributions of this paper are:
\begin{itemize}[leftmargin=*,nosep]
    \item To the best of our knowledge, we are the first to formulate and investigate the \acf{NNBR} task, which aims to recommend a set of novel items that meets a user's preferences in the next basket.
    \item We investigate the performance of several representative \acs{NBR} methods w.r.t.\ the \acs{NNBR} task and find
    \begin{enumerate*}[label=(\roman*)] \item that training specifically for the exploration task does not always lead to better performance, and \item that limited progress has been made w.r.t. the \acs{NNBR} task.\end{enumerate*}
    \item We propose a simple \acf{BTBR} model that learns item-to-item correlations across baskets.
    \item We propose several types of masking and item swapping strategies for optimizing \ac{BTBR} for the \acs{NNBR} task. Extensive experiments are done on three open grocery shopping datasets to assess the effectiveness of the proposed strategies. \ac{BTBR} with a proper masking and swapping strategy is the new \acl{SOTA} method w.r.t.\ the \acs{NNBR} task.
\end{itemize}


\section{Related Work}

In this section, we describe two lines of research in the recommender systems literature that are related to our work: sequential recommendation and next basket recommendation.

\header{Sequential recommendation}
Sequential item recommendation has been widely studied for many years, and models~\citep{gru4rec,gru4recupdate, stamp, caser,narm,srgnn,sasrec, bert4rec} with various deep learning techniques, e.g., RNN~\citep{gru4rec,gru4recupdate}, CNN~\citep{caser}, GNN~\citep{gnn, srgnn}, contrastive learning~\citep{xie2022contrastive}, attention~\citep{stamp, narm} and self-attention~\citep{vaswani2017attention,sasrec, bert4rec} mechanism have been proposed. 
The self-attention (transformer) model~\citep{vaswani2017attention} with multi-head attention shows strong performance in natural language processing, and SASRec~\citep{sasrec} is the first sequential recommendation model that employs the self-attention mechanism. 
BERT4Rec~\citep{bert4rec} upgrades the left-to-right training scheme in SASRec and uses a bi-direction transformer with a Cloze task~\citep{taylor1953cloze}, which is the closest sequential recommendation method to this paper.
Motivated by the success of BERT4Rec, some follow-up work has applied masked-item-prediction training to more specific scenarios~\citep{yu2022self}.

However, BERT4Rec and follow-up work only focus on the item sequential recommendation with only random masking during training~\citep{yu2022self}. We extend BERT4Rec to the basket sequence setting and propose several types of masking strategies and training objectives that are specifically designed for the \acs{NNBR} task.
Furthermore, in this work we study the next novel basket recommendation task, where both historical interactions and the predicted target are baskets (sets of items). None of the sequential recommendation models listed above have been designed to handle a sequence of baskets.

\header{Next basket recommendation}
Next basket recommendation is a sequential recommendation task that addresses the sequence of baskets in the grocery shopping scenario. Existing methods can be classified into three types: frequency neighbor-based methods~\citep{tifuknn,recency}, Markov chain (MC)-based methods~\citep{fpmc}, and deep learning-based methods~\citep{dream, beacon, clea, sets2sets, dnntsp, chen2021modelingcat,triple2vec,wang2019timestamp1,sun2020timestamp2, intnet, bai2018attribute, nbrdiversity, recanet, buycycle}.
Recently, \citet{nbr-rep-expl} have evaluated and assessed \acs{NBR} performance from a new repetition and exploration perspective; they find that the repetition task, i.e., recommending repeat items, is much easier than the exploration task, i.e., recommending explore items (a.k.a. novel items in this paper), besides the improvements of many recent methods come from the performance of the repetition task rather than better capturing correlations among items.
Inspired by this finding, an \acs{NBR} method~\citep{recanet} that only models the repetition behavior has been proposed, and an NBRR task~\citep{buycycle} that only focuses on recommending repeat items has been formulated. 

In this paper, we propose and formulate the \acl{NNBR} task that focuses on recommending novel items to the user, whereas all of the \acs{NBR} methods mentioned above focus on the conventional \acs{NBR} task, and their performance when generalized to the \acs{NNBR} task remains unknown.





\section{Task Formulation}


In this section, we describe and formalize the \acl{NNBR} task which is the focus of this paper. 

Formally, given a set of users $U = \{u_1$, $u_2$, \ldots, $u_n\}$ and a set of items $I = \{i_1 ,i_2, \ldots, i_m\}$, $S_u = [B_u^1, B_u^2, \ldots, B_u^t]$ represents the historical interaction sequence for user $u$, where $B_u^t$ represents a set of items $i\in{I}$ that user $u$ purchased at time step $t$. 
For a user $u$, the \emph{repeat item} $i_u^\mathit{rep}$ is the item that user $u$ has purchased before, which is defined as $i_u^\mathit{rep} \in I_{u}^\mathit{rep} = B_u^1 \cup B_u^2 \cup \ldots \cup B_u^t$, and the novel item $i_u^\mathit{novel}$ is the item that user $u$ have not purchased before, i.e., $i_u^\mathit{novel} \in I_u^\mathit{novel} = I - I_u^\mathit{rep}$.
 

The goal of the \emph{next novel basket recommendation} task is to predict the following novel basket which only consists of novel items $i_u^\mathit{novel}$ that the user would probably like, based on the user's past interactions $S_u$, that is,
\begin{equation}
P_{u} = \hat{B}_u^{t+1} = f(S_u)
\end{equation}
where $P_{u}$ denotes a recommended item list that \emph{only consists of the novel items $i_u^\mathit{novel}$ of user $u$}. 


\section{Our Method}
In this section, we first describe the base bi-directional transformer basket recommendation model (BTBR) we use, then introduce several types of masking strategies for the \acs{NNBR} task, and finally describe an item swapping strategy.
\subsection{Bi-directional transformer basket recommendation model}

Learning basket representations~\citep{basket-emb} and modeling temporal dependencies across baskets are two key components in almost all neural-based \acs{NBR} methods.
Many \acs{NBR} methods introduce complex architectures to learn representations for baskets  in grocery shopping~\citep{beacon, dnntsp, clea,sun2020timestamp2,chen2021modelingcat}.  
Instead of proposing more complex architectures to learn better basket representations and temporal dependencies, we want to simplify the model and only focus on the item-level correlations across different baskets, which helps us to infer novel items from users' historical items.

\begin{figure}
  \centering
  \includegraphics[width=0.5\linewidth]{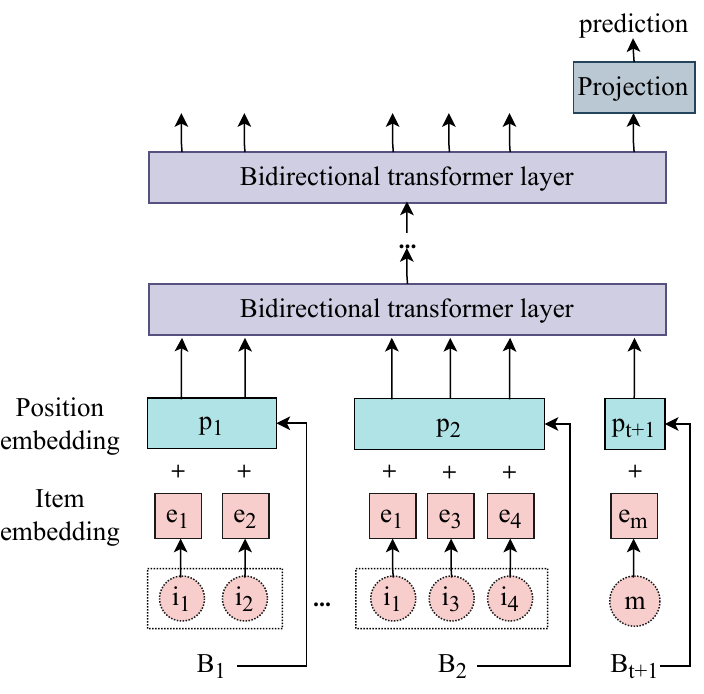}
  \caption{The overall architecture of the BTBR model.}
  \label{fig: btbr}
\end{figure}

As a widely used method to model temporal dependencies, a recurrent neural network (RNN)~\citep{lstm, gru} requires passing information sequentially according to the temporal order, whereas there is no temporal order for items within the same basket, and basket-level representations at each timestamp are required~\citep{dream, beacon, sets2sets, clea}. 
Another alternative method is the self-attention mechanism (a.k.a. transformer)~\citep{vaswani2017attention}, which is capable of learning the representations of every position by exchanging the information across all positions.
Therefore, we adopt the bi-directional transformer~\citep{vaswani2017attention,devlin2018bert} as the backbone of our BTBR model, which not only allows us to learn item-to-item correlations from the direct interactions among items across different baskets but is also able to handle basket sequence information in grocery shopping.
The overall architecture of BTBR is shown in Figure~\ref{fig: btbr}.


\header{Embedding layer}
In order to use transformers~\citep{vaswani2017attention} for \acs{NNBR}, we first transfer a basket sequence to an item sequence via a ``flatten'' operation, e.g., $[\{i_1, i_2\}, \{i_1, i_3, i_4\}] \rightarrow [i_1, i_2, i_1, i_3, i_4]$.
It has been shown that the positions of items are informative in the sequential recommendation scenario~\citep{sasrec,bert4rec}.
Different from solutions in conventional item sequential recommendation, where each item is combined with its unique position embedding w.r.t. its position in the item sequence, we use a learnable position embedding for every basket, and items within the same basket will share the same position embedding.
For example, given a basket sequence $S = [\{i_1, i_2\}, \{i_1, i_3, i_4\}, \{i_4, i_5\}]$, we first flatten $S$ and get a sequence of item embeddings $E_i = [e_1^i, e_2^i, e_1^i, e_3^i, e_4^i ,e_4^i ,e_5^i]$, and a position embedding sequence $E_p = [e_1^p, e_2^p, e_3^p]$. Finally, the input sequence of transformer layer will be $E_{i, p} = [e_1^i + e_1^p, e_2^i + e_1^p, e_1^i + e_2^p, e_3^i + e_2^p, e_4^i + e_2^p, e_4^i + e_3^p, e_5^i + e_3^p]$.
Note that the padding and truncating operations are also employed to handle sequences of various lengths.

\header{Bi-directional transformer layer}
The transformer architecture contains two sub-layers: 
\begin{enumerate}[leftmargin=*]
  \item \emph{Multi-head attention layer}, which adopts the popular attention mechanism~\citep{vaswani2017attention} and aggregates all items' embeddings across different baskets with adaptive weights.
  \item \emph{Point-wise feed-forward layer}, which aims to endow nonlinearity and interactions between different latent dimensions.
\end{enumerate}
We use stacked transformer layers to learn more complex item-to-item correlations, that is:
\begin{equation}
  H^1 = \operatorname{Trm}(E_{i, p}), \ldots, 
  H^L = \operatorname{Trm}(H^{L-1}),
\end{equation}
where $Trm$ denotes the bi-directional transformer layer, $H^L = [h_1^L, h_2^L, \dotsc, h_d^L]$ denotes a representation sequence derived from the last transformer layer, and $d$ denotes the maximum sequence length of input sequence $E_{i, p}$.
Besides, residual connections~\citep{residual}, dropout~\citep{dropout}, layer normalization~\citep{layernorm}, and GELU activation~\citep{gelu} are adopted to enhance the ability of representation learning. 
For more details about the bi-directional transformer layer, we refer to~\citep{sasrec, bert4rec,vaswani2017attention}.

\header{Prediction layer}
After hierarchically exchanging information of all items across baskets using the transformer, we get $H^L \in \mathbb{R}^{m\times d}$, which contains the corresponding representations $h^L$ for all items in the input sequence.
Following~\citep{sasrec,bert4rec}, we use the same item embedding $E_I\in\mathbb{R}^{m\times d}$ as the input layer to reduce the model size and alleviate the overfitting problem.
For a masked position (item), we get its learned representation $h\in \mathbb{R}^{d}$ and compute the interaction probability distribution $p$ of candidate items by:
\begin{equation}
  p = \operatorname{Softmax}(hE^\mathrm{T} + b),
  \label{eq:pred}
\end{equation}  
where $E$ is the embedding matrix for candidate items and $b$ denotes a bias term.

\subsection{Masking strategy}
\begin{figure}[t]
  \centering
  \includegraphics[width=0.5\columnwidth]{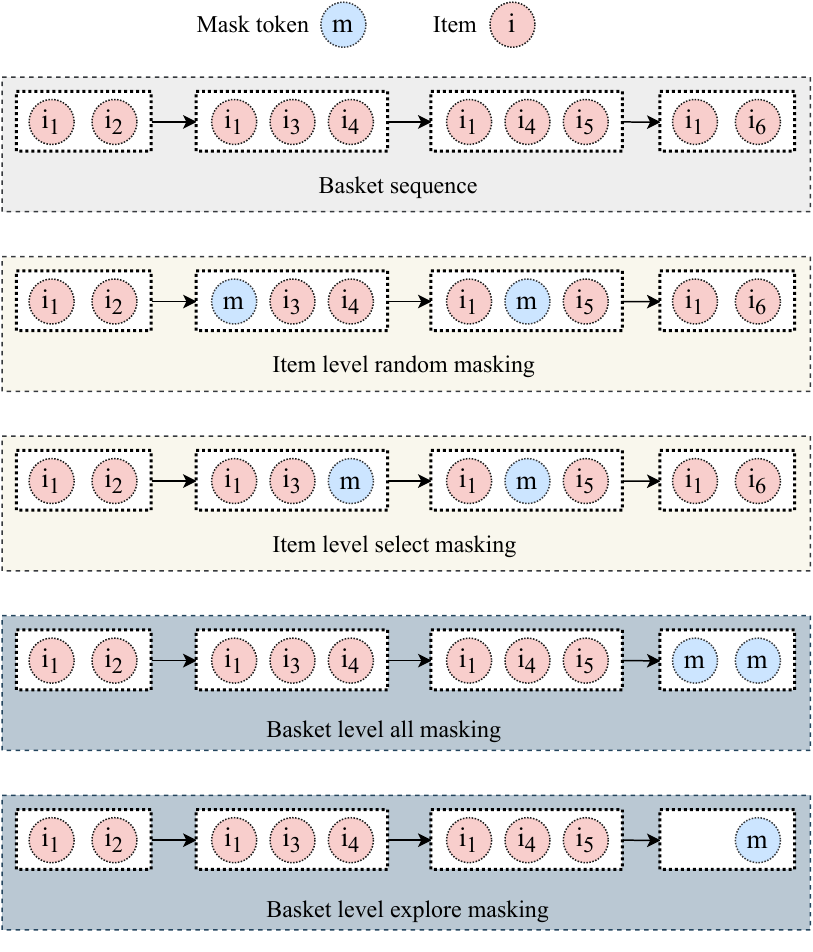}
  \caption{The original basket sequence (at the top) and four types of masking strategies.}
  \label{fig:masking}
\end{figure}

Since there are repetition signals in the basket sequence, it is unclear whether these signals are merely noise/shortcuts or contain valuable information for the task of recommending novel items.
After constructing the base model (BTBR), the challenging problem that needs to be addressed is how to properly train the model to improve its ability of finding novel items that meet users' interests.
In this section, we propose four types of alternative masking strategies for the \acl{NNBR} task by considering different tasks and levels, as well as the repetition-exploration signals.
Figure~\ref{fig:masking} shows examples of four types of masking strategies and Table~\ref{tab: strategy-difference} shows the characteristics of different training strategies.

\header{Cloze task} 
The first type of training objective is a cloze task~\citep{taylor1953cloze}, i.e., ``masked language model'' in~\citep{devlin2018bert}. 
Specifically, we mask a proportion of items in the input sequence, i.e., replace each of them with a ``mask token,'' and then try to predict the original items based on their contexts. 
We call this masking ``item-level.'' Two main advantages of this item-level masking \& training strategy are
\begin{enumerate*}[label=(\roman*)]
  \item it allows us to generate more item-level training samples by breaking the definition of ``basket,'' and
  \item it learns both sides' information via the bi-directional transformer, which might allow the model to better capture item-to-item correlations.
\end{enumerate*}
We first introduce two item-level masking strategies as follows:

\begin{enumerate}[leftmargin=*]
  \item \emph{Random}: This is a conventional masking strategy, which has been adopted in BERT4Rec~\citep{bert4rec}. Specifically, given a flattened item sequence, we randomly select several positions of the sequence and mask the corresponding items of the selected position according to mask ratio $\alpha$ as input.
  \item \emph{Select}: One potential issue w.r.t. the above \emph{Random} masking is that the masked items (prediction target) might still exist in the non-masked positions, so the model might mainly predict the masked item via its repetition information rather than inferring new items based on item-to-item correlations. Therefore, we propose the select masking strategy, which is specifically targeted at the exploration demand of the \acs{NNBR} task. 
  Specifically, given a flattened item sequence, we first derive the item set $I$ in this sequence, then randomly select several items $i_m \in I$ according to mask ratio $\alpha$, and finally mask all the occurrences of $i_m$ in the sequence. Since there is no repetition information available, the model can only infer the targeted items, i.e., novel items, via learning the item-to-item correlations.
\end{enumerate}

\header{Basket recommendation task}
Using the cloze task as the learning objective has limitations: 
\begin{enumerate*}[label=(\roman*)]
\item it is not able to fully respect the temporal dependencies of a sequence, since we can only use the historical information (left-side context) when we make the recommendation;  and
\item it is not specifically designed for the basket recommendation task and a mismatch might exist.
\end{enumerate*} 
Therefore, the second type of training objective we consider is the basket recommendation task, which masks the input sequence at the basket-level instead of item-level. Specifically, we mask the last basket and try to predict the items in this basket only based on the historical items (left-side information).
Similarly, we propose another two basket-level masking strategies as follows:
\begin{enumerate}[leftmargin=*,resume]
\item \emph{All}: This masking strategy can be regarded as optimizing the model for the \acs{NBR} task. Given a flattened item sequence, we find and mask all items, i.e., both novel items and repeat items in the last basket. 
\item \emph{Explore}: This is a \acs{NNBR}-specific masking strategy. Given a flattened item sequence, we find the items in the last basket, instead of masking all items, we only mask the novel items $i\in I^\mathit{novel}$ and remove the \emph{repeat items} $i\in I^\mathit{rep}$. The model will be only optimized for finding all novel items in the future based on the historical basket sequence. 
\end{enumerate}

\header{Joint task}
The pretrain-then-finetune paradigm has been widely adopted in NLP tasks.
It is worth noting that item-level masking (the cloze task) and basket-level masking (the basket recommendation task) can also be combined as a joint masking strategy to employ the pretrain-then-finetune paradigm in \acs{NNBR}, which first uses item-level masking strategy (i.e., self-supervised task) to get item correlations as the pre-train stage and then employ basket-level masking strategy (i.e., supervised task) to finetune it for the basket recommendation task. 

\begin{table}
  \newcommand{\tabincell}[2]{\begin{tabular}{@{}#1@{}}#2\end{tabular}}
    \caption{Comparison of four types of masking strategies from three aspects, i.e., temporal orders, explore specific and amount of training signals.}
    \label{tab: strategy-difference}
    \setlength{\tabcolsep}{2pt}
    \begin{tabular}{l@{}cccc}
      \toprule
      \bf Strategy & Strict temporal orders & Explore specific& Training signals ranking\tablefootnote{As item-level masking can be seen as self-supervised learning, which is more flexible and can leverage more training signals than basket-level masking. Basket-explore has the least training signals as it can only use the novel items in the last basket.}\\
      \midrule
      Item-Random & $\times$ & $\times$& 1\\
      Item-Explore & $\times$  & $\checkmark$& 1\\
      Basket-All & $\checkmark$ & $\times$& 2 \\
      Basket-Explore&$\checkmark$&$\checkmark$& 3\\
      \bottomrule
    \end{tabular}
\end{table}

\header{Loss}
Following~\citep{bert4rec}, we select minimizing the negative log-likelihood loss as the training objective:
\begin{equation}
  \mathcal{L} = \frac{1}{|I^m|}\sum_{i\in I^m}-\log p(i \mid S_u),
\end{equation}
where $I^m$ is the masked item set, $p(i\mid S_u, t)$ is the predicted probability of item $i$ at position $t$.

\header{Test and prediction}
To predict a future basket (a set of items), we only need to add one masked token at the end of the user's item sequence, since items within the same basket share the same position embedding.
In the \acs{NNBR} task, the candidate items are novel items $I^\mathit{new}$ that the user has not bought before, thus we use the embedding matrix w.r.t.\ the novel items of every user to compute the probabilities according to Eq.~\ref{eq:pred}.
Finally, we select top-$K$ novel items with the highest scores as the recommendation list of the next novel basket.

\subsection{Swapping strategy}

In sequential recommendation, some work~\citep{flexible-order, cheng2021learning, xie2022contrastive, sequential-survey} argues that the items in a sequence may not be sequentially dependent and different item orders may actually correspond to the same user intent.
Ignoring flexible orders in sequential recommendation might lead to less accurate recommendations for scenarios where many items are not sequentially dependent~\citep{sequential-survey, yu2022self, flexible-order}. 
In grocery shopping, the items purchased within the different baskets might not have rigid orders. 
To further understand if considering the flexible orders among items could further improve the performance w.r.t. the \acs{NNBR} task, we propose the item swapping strategy to create augmentations for the BTBR.

\begin{figure}[h]
  \centering
  \includegraphics[width=0.5\linewidth]{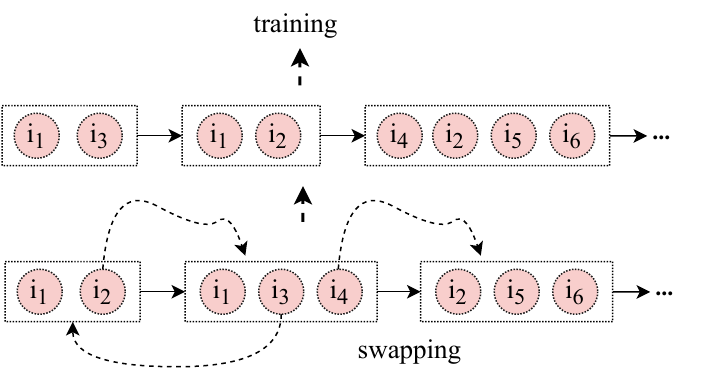}
  \caption{An example of the item swapping strategy.}
  \label{fig:swapping}
\end{figure}

Specifically, as illustrated in Figure~\ref{fig:swapping}, we randomly select items according to a swap ratio $\lambda$ and then move them to another basket to enrich the items' interactions within the same basket. Besides, we introduce a hyper-parameter, i.e., swap hop $\gamma$, to control the basket distance of the swapping strategy.
Note that we only perform the local swap strategy when using item-level masking (the cloze task) to train the model, since basket-level masking (the basket recommendation task) is designed to respect the sequential order and predict the future basket based on historical information.

\section{Experiments}

\subsection{Research questions}
To understand the \acl{NNBR} task, and evaluate the performance of BTBR with different strategies, we conduct experiments to answer the following questions:
\begin{enumerate}[label=\textbf{RQ}\arabic*,leftmargin=*]
\item How do existing \acs{NBR} models perform w.r.t. the \acs{NNBR} task? Does training specifically for the \acs{NNBR} task lead to better performance?

\item How does BTBR with different masking strategies perform compared to the state-of-the-art models? 

\item Does the swapping strategy contribute to the improvements? 

\item How do the hyper-parameters influence the models' performance and how different masking strategies affect the training dynamics? 

\item Is the joint masking strategy more robust than using the single masking strategy?
\end{enumerate}

\subsection{Experimental setup}
\textbf{Datasets.}
We evaluate the \acs{NNBR} task on three publicly available grocery shopping datasets (TaFeng,\footnote{\url{https://www.kaggle.com/chiranjivdas09/ta-feng-grocery-dataset}} Dunnhumby,\footnote{\url{https://www.dunnhumby.com/source-files/}}  and Instacart\footnote{\url{https://www.kaggle.com/c/instacart-market-basket-analysis/data}}), which vary in their repetition and exploration ratios.
Following~\cite{nbr-rep-expl}, we sample users whose basket length is between 3 and 50, and remove the least frequent items in each dataset. We also focus on the fixed size (10 or 20) next novel basket recommendation problem.
In our experiments, we split the dataset across users, 80\% for training, and 20\% for testing, and leave 10\% of the training users as the validation set.
We repeat the splitting and experiments five times and report the average performance.
The statistics of the processed datasets are shown in Table~\ref{dataset}.

\begin{table}[h]
\newcommand{\tabincell}[2]{\begin{tabular}{@{}#1@{}}#2\end{tabular}}
  \caption{Statistics of the processed datasets. }
  \label{dataset}
  \setlength{\tabcolsep}{2pt}
  \begin{tabular}{l@{}cccccc}
    \toprule
    \bf Dataset & \bf \#items & \bf \#users & \tabincell{c}{\bf Avg. \\ \bf basket\\\bf size} & \tabincell{c}{\bf Avg. \\ \bf \#baskets\\\bf per user}  & \tabincell{c}{\bf repeat\\\bf ratio} & \tabincell{c}{\bf explore\\\bf ratio}\\
    \midrule
    TaFeng & 11,997 & 13,858 & 6.27 & \phantom{1}6.58 & 0.188 & 0.812 \\
    Dunnhumby & \phantom{0}3,920 & 22,530 & 7.45 & \phantom{1}9.53 & 0.409 & 0.591\\
    Instacart & 13,897 & 19,435 & 9.61 & 13.21 & 0.597 & 0.403\\
    \bottomrule
  \end{tabular}
\end{table}

\header{Baselines}
We investigate the performance of six \acs{NBR} baselines, which we select based on their performance on our chosen datasets in the analysis performed in~\cite{nbr-rep-expl}. Importantly, for a fair comparison, we do not include methods that leverage additional information~\citep{bai2018attribute,sun2020timestamp2,chen2021modelingcat}.

\iftrue
\begin{itemize}[leftmargin=*,nosep]
  \item \textbf{G-TopFreq:} A simple and effective method that recommends the top $k$ most popular items in the dataset as the next basket for users.
  \item \textbf{TIFUKNN:} A state-of-the-art method that models the temporal dynamics of users' past baskets by using a KNN-based approach based on personalized frequency information (PIF)~\citep{tifuknn}.
  \item \textbf{Dream:} A RNN-based method that gets basket representation using pooling strategy and employs RNN to model sequential behavior~\citep{dream}.
  \item \textbf{Beacon:} A RNN-based method that uses RNN to capture sequential behavior and uses correlation-sensitive basket encoder to consider intra-basket item correlations~\citep{beacon}.
  \item \textbf{DNNTSP:} A state-of-the-art method that utilizes a graph neural network (GNN) and self-attention mechanisms to encode item-item relations across baskets and capture temporal dependencies~\citep{dnntsp}.
  \item \textbf{CLEA:} A state-of-the-art method that uses contrastive learning and a GRU-based encoder to denoise and automatically extract items relevant to the target item~\citep{clea}.
\end{itemize}
\fi

\noindent%
Note that for the above baseline models (except G-TopFreq), we have two versions with different training methods, i.e., using all items in the last basket as training labels (Train-all), and only using novel items in the last basket as training labels (Train-explore). 

\header{Configurations}
For the training-based baseline methods and TIFUKNN, we strictly follow the hyper-parameter setting and tuning strategy of their respective original papers. The embedding size is tuned on $\{16, 32, 64, 128\}$ for all training-based methods based on the validation set to achieve their best performance. 


We use PyTorch to implement our model and train it using a TITAN X GPU with 12G memory. For BTBR, we set self-attention layers to 2 and their head number to 8, and tune the embedding size on $\{16, 32, 64, 128\}$.
The Adam optimizer with a learning rate of 0.001 is used to update parameters. We set the batch size to 128 for the Tafeng and Dunnhumby datasets, and 64 for the Instacart dataset; we sweep the mask ratio $\alpha$ in $\{0.1, 0.3, 0.5, 0.7, 0.9\}$, local swap ratio in $\{0, 0.1, 0.3, 0.5, 0.7, 0.9\}$ and swap hop $\gamma$ in $\{1, 3, 5, 7, 9\}$.

\header{Metrics}
Two widely used metrics for the \acs{NBR} problem are $Recall@k$ and $nDCG@k$. In the \acs{NNBR} task, $Recall$ measures the ability to find all novel items that a user will purchase in the next basket; NDCG is a ranking metric that also considers the order of these novel items, i.e., 
\begin{align}
  \mathrm{Recall}@K &  = \frac{1}{|U|}\sum_{u\in U}\frac{\left| P_{u} \cap T_{u}^\mathit{novel}\right|}{\left| T_{u}^\mathit{novel}\right|},
\\
  \mathrm{nDCG}@K & = \frac{1}{|U|}\sum_{u\in U}\frac{\sum_{k=1}^K p_k/\log_2(k+1)}{\sum_{k=1}^{\min(K, |T_{u}^\mathit{novel}|)} 1/\log_2(k+1)},
\end{align}
where $U$ is a set of users who will purchase novel items in their next basket, $T_{u}^\mathit{novel}$ is a set of ground-truth novel items of user $u$, $p_k$ equals 1 if $P_{u}^k\in T_u^{novel}$, otherwise $p_k=0$.  $P_u^k$ denotes the $k$-th item in the predicted basket $P_u$.
Note that some methods might assign high scores w.r.t. the repeat items~\citep{nbr-rep-expl}, to generate a novel basket, we fully remove the repeat items, then only rank and select top-$k$ novel items as the recommended basket $P_u$ to ensure a fair comparison, i.e., the recommended basket \emph{only consists top-$k$ novel items}.

\begin{table*}
	\centering
	\caption{Results of methods training for finding novel items, i.e., Train-explore, compared against the methods training for finding all items, i.e., Train-all. Boldface and underline indicate the best and the second best performing performance w.r.t. the \acs{NNBR} task, respectively. Significant improvements and deteriorations of Train-explore over the corresponding Train-all baseline results are marked with~$^\uparrow$ and $^\downarrow$, respectively.~(paired t-test, p $<$ 0.05).  }
	
	\begin{tabular}{@{} l l l l l l l l l }
		\toprule
		Dataset& Metric& Train&G-Pop& TIFUKNN &Dream &Beacon & CLEA& DNNTSP\\
		\midrule
		
		\multirow{8}{*}{\rotatebox[origin=c]{90}{Tafeng}}
    & \multirow{2}{*}{Recall@10}&all&0.0587&0.0714&0.0960&0.0926&0.0870&\bf{0.1024}\\
		&&explore&=&0.0911$^\uparrow$&\underline{0.1021}$^\uparrow$&0.0967$^\uparrow$&0.1010$^\uparrow$&0.0940$^\downarrow$\\
    \cmidrule{2-9}
    & \multirow{2}{*}{nDCG@10}&all&0.0603&0.0662&0.0823&0.0789&0.0755&\underline{0.0855}\\
    &&explore&=&0.0783$^\uparrow$&\bf{0.0859}$^\uparrow$&0.0819$^\uparrow$&0.0857$^\uparrow$&0.0767$^\downarrow$\\
    \cmidrule{2-9}
    & \multirow{2}{*}{Recall@20}&all&0.0874&0.0926&0.1244&0.1252&0.1150&0.1245\\
    &&explore&=&0.1157$^\uparrow$&0.1244&\bf{0.1257}&\underline{0.1253}$^\uparrow$&0.1168$^\downarrow$\\
    \cmidrule{2-9}
    & \multirow{2}{*}{nDCG@20}&all&0.0703&0.0738&0.0928&0.0909&0.0861&\underline{0.0943}\\
		&&explore&=&0.0876$^\uparrow$&0.0939&0.0929&\bf{0.0952}$^\uparrow$&0.0858$^\downarrow$\\
		\cmidrule{1-9}
		
		\multirow{8}{*}{\rotatebox[origin=c]{90}{Dunnhumby}}
    & \multirow{2}{*}{Recall@10}&all&0.0468&0.0497&0.0494&0.0499&0.0499&0.0514\\
    &&explore&=&0.0498&0.0506&\bf{0.0529}$^\uparrow$&\underline{0.0520}$^\uparrow$&0.0472$^\downarrow$\\
    \cmidrule{2-9}
    & \multirow{2}{*}{nDCG@10}&all&0.0397&0.0409&0.0409&0.0411&0.0376&\underline{0.0415}\\
    &&explore&=&0.0411&0.0385&\bf{0.0428}$^\uparrow$&0.0404$^\uparrow$&0.0378$^\downarrow$\\
    \cmidrule{2-9}
    & \multirow{2}{*}{Recall@20}&all&0.0701&0.0745&0.0744&0.0804&0.0711&0.0782\\
    &&explore&=&0.0746&0.0791&\bf{0.0813}&\underline{0.0807}$^\uparrow$&0.0739\\
    \cmidrule{2-9}
    & \multirow{2}{*}{nDCG@20}&all&0.0491&0.0505&0.0505&\underline{0.0532}&0.0479&0.0524\\
    &&explore&=&0.0506&0.0502&\bf{0.0546}&0.0521$^\uparrow$&0.0484$^\downarrow$\\
		\cmidrule{1-9}
		
		\multirow{8}{*}{\rotatebox[origin=c]{90}{Instacart}}
    & \multirow{2}{*}{Recall@10}&all &0.0430&0.0425&0.0440&0.0454&0.0394&0.0414\\
    &&explore&=&\bf{0.0494}$^\uparrow$&0.0455&0.0460&\underline{0.0469}$^\uparrow$&0.0419\\
    \cmidrule{2-9}
    & \multirow{2}{*}{nDCG@10}&all&0.0359&0.0346&0.0356&0.0388&0.0302&0.0335\\
    &&explore&=&\bf{0.0400}$^\uparrow$&0.0355&\underline{0.0387}&0.0369$^\uparrow$&0.0341\\
    \cmidrule{2-9}
    & \multirow{2}{*}{Recall@20}&all&0.0685&0.0649&0.0690&0.0733&0.0626&0.0635\\
    &&explore&=&\underline{0.0755}$^\uparrow$&0.0719&0.0741&\bf{0.0764}$^\uparrow$&0.0642\\
    \cmidrule{2-9}
    & \multirow{2}{*}{nDCG@20}&all& 0.0455&0.0431&0.0452&0.0499&0.0394&0.0424\\
    &&explore&=&\underline{0.0500}$^\uparrow$&0.0462&\bf{0.0501}&0.0484$^\uparrow$&0.0431\\
		\bottomrule			
	\end{tabular}	
	\label{tab:nbr_results}

\end{table*}

\subsection{Train-all and Train-explore (RQ1)}
\label{sec: nbr}

To answer RQ1, we employ two training strategies for each baseline method: 
\begin{enumerate*}[label=(\roman*)]
\item \emph{Train-all}: we keep both repeat items and explore items as part of the ground-truth labels during training, which means that the model is trained to find all possible items in the next basket; and
\item \emph{Train-explore}: we remove the repeat items and only keep novel items in the ground-truth labels during training, which means the model is specifically trained to find novel items in the next basket.
\end{enumerate*}
For the NNBR performance evaluation, we assess the models' ability to find novel items, which means the recommended novel basket consists of top-$k$ novel items and there are no repeat items. We report the experimental results of different baseline methods in Table~\ref{tab:nbr_results}.
We have three main findings.

First, we see that no method consistently outperforms all other methods across all datasets. 
On the Tafeng dataset, several NN-based methods (Dream-all, Dream-explore, Beacon-all, Beacon-explore, DNNTSP-all, CLEA-explore) fall in the top-tier methods group with quite good performance. 
On the Dunnhumby dataset, Beacon-explore achieves the best performance w.r.t.\ all metrics. On the Instacart dataset, TIFUKNN-explore is among the best-performing methods, which means that well-tuned neighbor-based models may outperform complex neural-based methods on some datasets w.r.t. the \acs{NNBR} task~\citep{realprogress,nbr-rep-expl}.
The performance of G-TopFreq is obviously the worst on the Tafeng and Dunnhumby dataset, however, its performance is quite competitive on the Instacart dataset, which indicates that the popularity information is very important w.r.t. the \acs{NNBR} task in the scenario with a high repeat ratio. 

Second, the improvements of recent methods achieved in \acs{NBR} task do not always generalize to the \acs{NNBR} task. Recent proposed methods (TIFUKNN, CLEA, DNNTSP) have surpassed the previous classic baselines (i.e., G-TopFreq, Dream, Beacon) by a large margin in conventional \acs{NBR} task~\citep{clea,dnntsp,tifuknn,nbr-rep-expl}, whereas, the improvements are relatively small or even missing on some datasets when handling the \acs{NNBR} task. This indicates that the recently proposed methods make limited progress on finding novel items for the user and that their improvements mainly come from the repeat recommendation, which is consistent with the findings in~\citep{nbr-rep-expl}.

Third, the \acs{NNBR} performance changes diversely for different methods when changing from Train-all to Train-explore. Training and tuning existing \acs{NBR} methods specifically for the \acs{NNBR} task lead to significant or mild improvements in most cases, since the models do not need to deal with the repetition task and they are more targeted on finding novel items that meet users' preferences. 
Surprisingly, we find that DNNTSP-explore's performance is much worse than DNNTSP-all on the Tafeng and Dunnhumby datasets. We suspect that the underlying reason for this deterioration is that the repeat items (labels) contain useful item-to-item correlation signals that can be captured by the DNNTSP.\footnote{Assume that one user's historical basket sequence is $[[a, b, c], [c, d], [a, c]]$, and next basket is $[b, e]$. Even though $b$ is a repeat item, the model might be able to learn the correlation between $b$ and other items in this historical sequence, which might help with the model's ability of finding novel items.}
Since various NBR methods have distinct architectures, certain methods may gain more from tailored training for exploration, while others can grasp item-item correlations from repeat labels. Consequently, it is unwise to indiscriminately eliminate repeat labels during training.\footnote{This finding is important as it helps to avoid the potential issue of poor baselines. To ensure a fair comparison, \acs{NNBR} practitioners should experiment with both strategies to train their baseline models and achieve best performances, instead of using an intuitive solution, i.e., removing repeat labels. }


\subsection{Effectiveness of BTBR (RQ2)}
\label{sec: nnbr}

\begin{table*}
	\centering
	\caption{Results of BTBR method with different masking strategies compared against the best performance of baseline method training for each metric w.r.t. \acs{NNBR} task. Boldface and underline indicate the best and the second best performing performance w.r.t. the \acs{NNBR} task, respectively. Significant improvements and deteriorations of over the best baseline results are marked with~$^\uparrow$ and $^\downarrow$, respectively (paired t-test, p $<$ 0.05). $\blacktriangle \%$ shows the improvements against the best performing baseline.}
	
	\begin{tabular}{@{} l l c @{\quad} c c c c @{\quad} c c @{\quad} c }
		\toprule
		\multirow{3}{*}{\rotatebox[origin=c]{90}{Dataset}} & & & \multicolumn{4}{c}{Item level} & \multicolumn{2}{c}{Basket level} & Joint\\
		\cmidrule(l{0pt}r{6pt}){4-7}
    \cmidrule(l{0pt}r{6pt}){8-9}
		\cmidrule(r{0pt}){10-10}
		& Metric& Best & Random& Select&
    \vtop{\hbox{\strut Random}\hbox{\strut swap}}& \vtop{\hbox{\strut Select}\hbox{\strut swap}}&
     All& Explore&\vtop{\hbox{\strut Pretrain-Finetune}}\\
		\midrule
		
		\multirow{4}{*}{\rotatebox[origin=c]{90}{Tafeng}}
    & Recall@10&0.1024&0.0736$^\downarrow$&0.0801$^\downarrow$&0.0717$^\downarrow$&0.0746$^\downarrow$&\underline{0.1056}$^\uparrow$&0.1032&\bf{0.1057}$^\uparrow$\small{(3.2\%)}\\
		& nDCG@10&0.0859&0.0597$^\downarrow$&0.0651$^\downarrow$&0.0587$^\downarrow$&0.0605$^\downarrow$&\underline{0.0869}&0.0860&\bf{0.0870}\phantom{$^\uparrow$}\small{(1.3\%)}\\
		& Recall@20&0.1257&0.0977$^\downarrow$&0.1036$^\downarrow$&0.0895$^\downarrow$&0.0911$^\downarrow$&\underline{0.1292}$^\uparrow$&0.1271&\bf{0.1353}$^\uparrow$\small(7.6\%)\\
		& nDCG@20&0.0952&0.0691$^\downarrow$&0.0739$^\downarrow$&0.0685$^\downarrow$&0.0688$^\downarrow$&\underline{0.0970}$^\uparrow$&0.0957&\bf{0.0973}$^\uparrow$\small(2.2\%)\\
		
		\cmidrule{2-10}
		
		\multirow{4}{*}{\rotatebox[origin=c]{90}{Dunnhumby}}
    & Recall@10&0.0529&0.0548$^\uparrow$&0.0572$^\uparrow$&0.0553$^\uparrow$&\underline{0.0592}$^\uparrow$&0.0524&0.0521&\bf{0.0593}$^\uparrow$\small(12.1\%)\\
		& nDCG@10&0.0428&0.0439$^\uparrow$&0.0461$^\uparrow$&0.0443$^\uparrow$&\bf{0.0469}$^\uparrow$&0.0427&0.0424&\underline{0.0468}$^\uparrow$\small(9.3\%)\\
		& Recall@20&0.0813&0.0847$^\uparrow$&0.0891$^\uparrow$&0.0867$^\uparrow$&\bf{0.0924}$^\uparrow$&0.0815&0.0806&\underline{0.0915}$^\uparrow$\small(12.5\%)\\
		& nDCG@20&0.0546&0.0560$^\uparrow$&0.0587$^\uparrow$&0.0571$^\uparrow$&\bf{0.0598}$^\uparrow$&0.0540&0.0532&\underline{0.0596}$^\uparrow$\small(9.2\%)\\
		
		\cmidrule{2-10}
		
		\multirow{4}{*}{\rotatebox[origin=c]{90}{Instacart}}
    & Recall@10&0.0494&0.0554$^\uparrow$&0.0583$^\uparrow$&0.0572$^\uparrow$&\bf{0.0600}$^\uparrow$&0.0539$^\uparrow$&0.0455$^\downarrow$&\underline{0.0598}$^\uparrow$\small(21.1\%)\\
		& nDCG@10&0.0400&0.0445$^\uparrow$&0.0474$^\uparrow$&0.0458$^\uparrow$&\bf{0.0486}$^\uparrow$&0.0426$^\uparrow$&0.0387$^\downarrow$&\underline{0.0478}$^\uparrow$\small(19.5\%)\\
		& Recall@20&0.0764&0.0887$^\uparrow$&0.0924$^\uparrow$&0.0898$^\uparrow$&\bf{0.0935}$^\uparrow$&0.0846$^\uparrow$&0.0734$^\downarrow$&\underline{0.0934}$^\uparrow$\small(22.3\%)\\
		& nDCG@20&0.0501&0.0573$^\uparrow$&0.0607$^\uparrow$&0.0583$^\uparrow$&\bf{0.0616}$^\uparrow$&0.0551$^\uparrow$&0.0485$^\downarrow$&\underline{0.0613}$^\uparrow$\small(22.4\%)\\

		\bottomrule			
	\end{tabular}	
	\label{tab:nnbr_results}

\end{table*}

In this experiment, we evaluate the overall \acs{NNBR} task performance of BTBR with different masking strategies, i.e., item-level random masking (item-random), item-level select masking (item-select), basket-level all masking (basket-all) and basket-level explore masking (basket-explore).
The results of the comparison with the best baseline performances are shown in Table~\ref{tab:nnbr_results}.\footnote{To avoid confusion, we only mark the significant differences for comparison with the baselines in this table. More comparison results among different strategies can be found in the experimental analysis.} Based on the results, we have several observations.
First, BTBR with the basket-all masking strategy (i.e., conventional next basket recommendation task) can significantly outperform the best baselines on the Tafeng and Instacart datasets, and achieve comparable performance on the Dunnhumby dataset.
This result indicates that it may not be necessary to introduce basket representations, because only modeling item-to-item correlations is already effective for the \acs{NNBR} task.

Second, there is no consistent best masking strategy across all datasets. 
On the Tafeng dataset, it is clear that basket-level masking outperforms item-level masking, where basket-all and basket-explore can respectively outperform and achieve the existing best performances w.r.t. each metric; however, using item-level masking leads to significant deterioration.  
On the Dunnhumby and Instacart datasets, BTBR with item-level masking strategies significantly outperforms the best performance achieved by baselines by a large margin, and is superior to BTBR with basket-level masking strategies.
The above results show that the sequential order of items or baskets on the Tafeng dataset might be more strict than the order on the Dunnhumby and Instacart datasets, so using item-level masking, which fails to fully respect the sequential order and has poor performance on the Tafeng dataset.

Third, we can also observe that item-select masking achieves better performance than item-random masking w.r.t. all metrics across all datasets (paired t-test, $p<0.05$), i.e., the improvements range from 4.1\% to 9.0\%, which demonstrates the effectiveness of our specifically designed item-select masking strategy for the \acs{NNBR} task.
In a sequence with many recurring items, the conventional random masking strategy could not ensure there is no masked item remaining in the other positions of the sequence, so the model might learn to predict the masked item based on the items' remaining occurrences, i.e., item self-relations. While the proposed item-select masking strategy will remove all occurrences of the same item, which can ensure that the masked items are novel items w.r.t. the remaining masked sequence, and the model has to infer the masked novel item via learning the masked item's relation with other items.

Finally, it can also be seen that basket-explore masking, which is specifically targeted at the \acs{NNBR} task, does not lead to any improvements on the Tafeng and Dunnhumby datasets, and results in a decrease in performance on the Instacart dataset, compared with basket-all masking. This result again verifies the findings in Section~\ref{sec: nbr} and indicates that masking and training BTBR specifically for the NNBR task may be suboptimal, since the repeat item labels may also be helpful with item-to-item correlations modeling. 

\subsection{Effectiveness of the item swapping strategy (RQ3)}
In this section, we conduct experiments to verify the effectiveness of the swapping strategy, and the results are shown in Table~\ref{tab:nnbr_results}.   
We find that adding a swapping strategy on top of item-random and item-select leads to a decrease in performance on the Tafeng dataset. 
At the same time, we note that adding a swapping strategy on top of item-random and item-select leads to better performance on the Dunnhumby and Instacart datasets (paired t-test, $p<0.05$).
These results are not surprising, since the swapping strategy will not only enrich the item interactions within the basket, but also has a risk of introducing noise w.r.t. the temporal information. 
The sequential order is relatively strict on the TaFeng dataset (see Section~\ref{sec: nnbr}), and the model can not benefit from the swap strategy.

\begin{figure}
    \centering
    \begin{subfigure}[b]{0.3\linewidth}
      \flushleft
      \includegraphics[width=\linewidth]{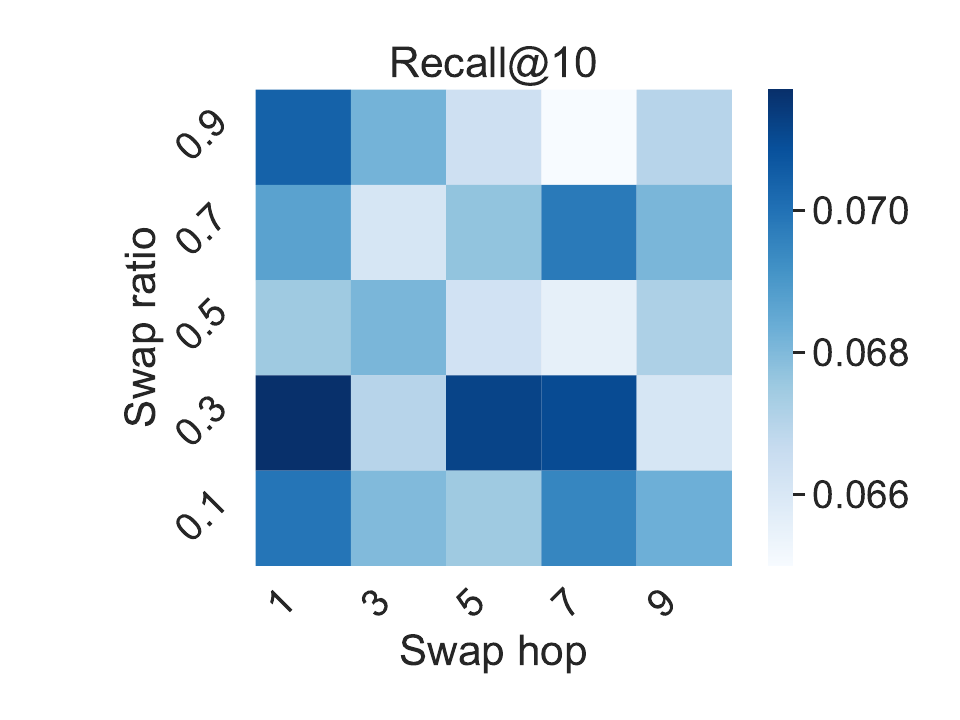}
      \vspace*{-7mm}
      \caption{{\small Tafeng (item-random)}}    
  \end{subfigure}
  \hspace{-2mm}
  \begin{subfigure}[b]{0.3\linewidth}  
      \flushright
      \includegraphics[width=\linewidth]{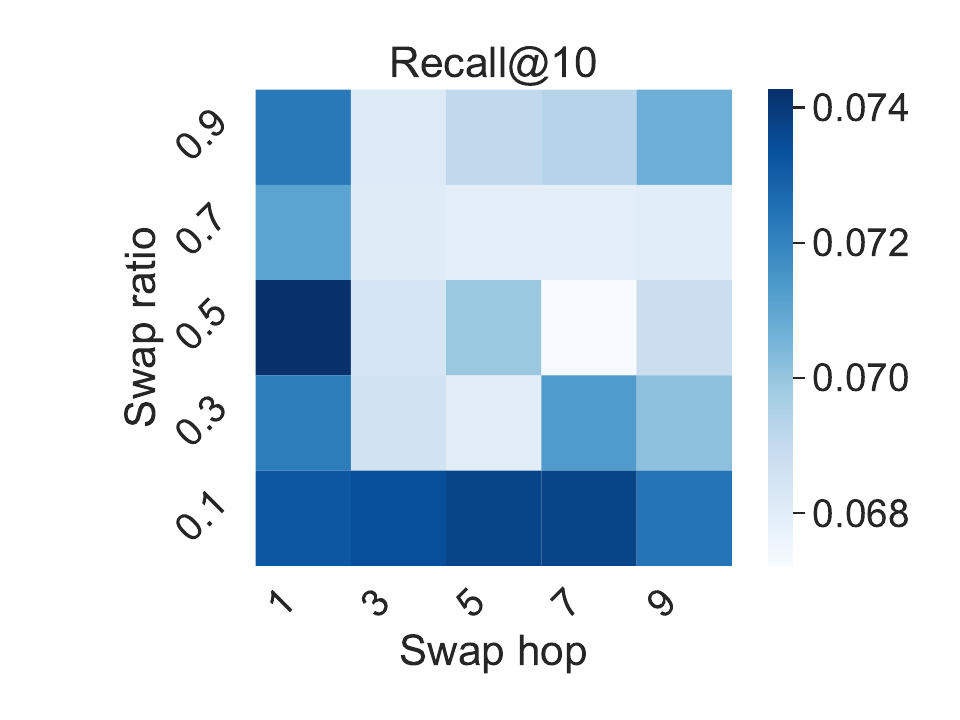}
      \vspace*{-7mm}
      \caption{{\small Tafeng (item-select)}}    
  \end{subfigure}
    \begin{subfigure}[b]{0.3\linewidth}
        \centering
        \includegraphics[width=\linewidth]{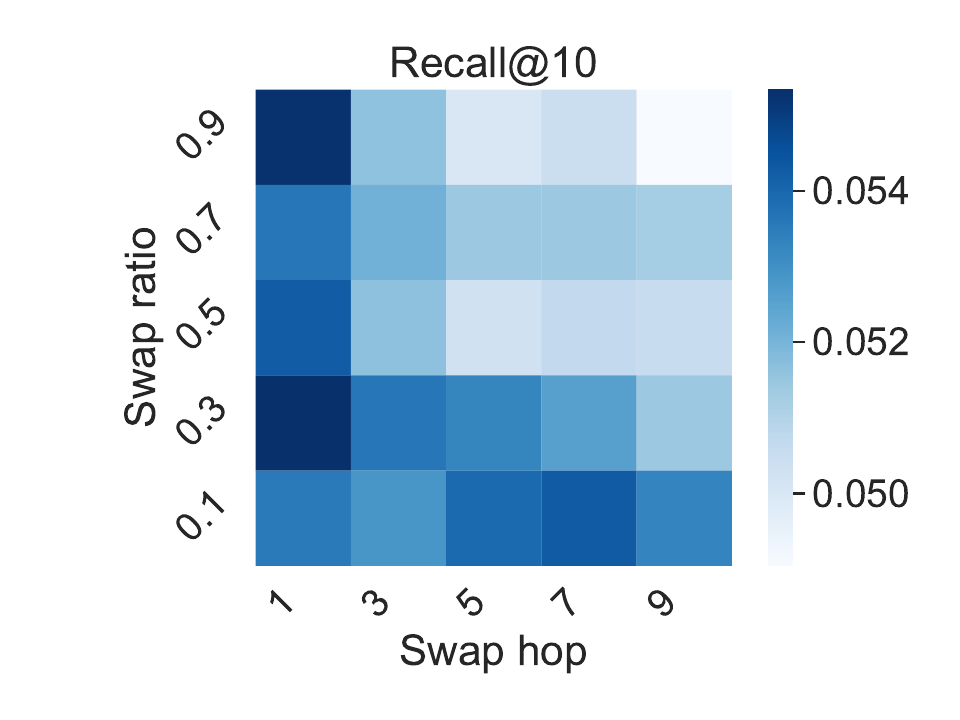}
      \vspace*{-7mm}
      \caption{{\small Dunnhumby (item-random)}}    
    \end{subfigure}
    \hspace{-2mm}
    \begin{subfigure}[b]{0.3\linewidth}  
        \centering 
        \includegraphics[width=\linewidth]{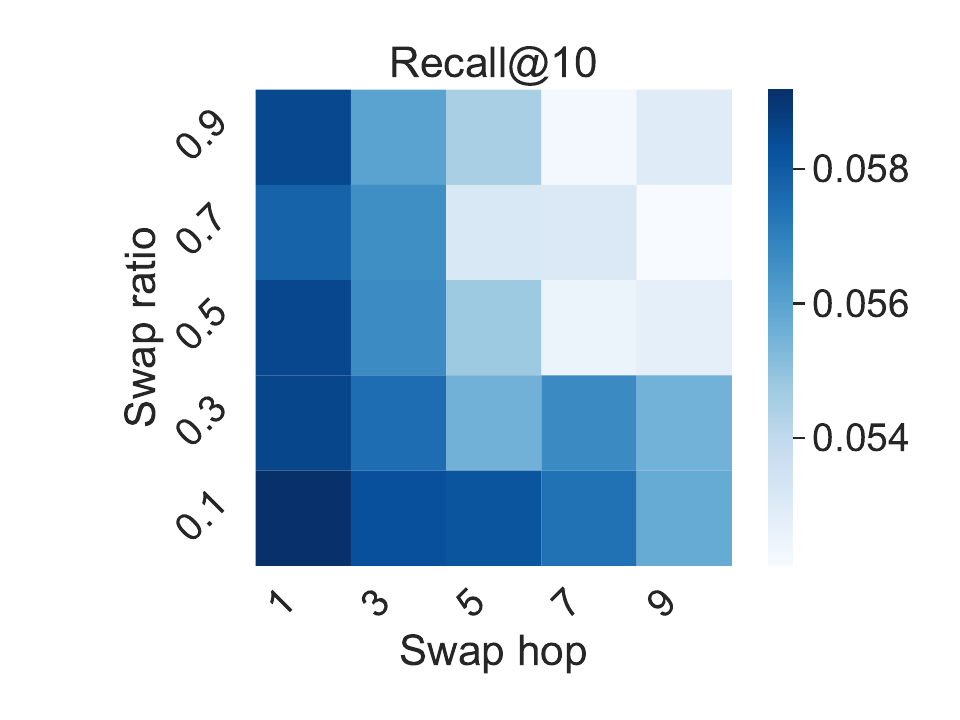}
      \vspace*{-7mm}
      \caption{{\small Dunnhumby (item-select)}}       
    \end{subfigure}
    \begin{subfigure}[b]{0.3\linewidth}   
        \centering 
        \includegraphics[width=\linewidth]{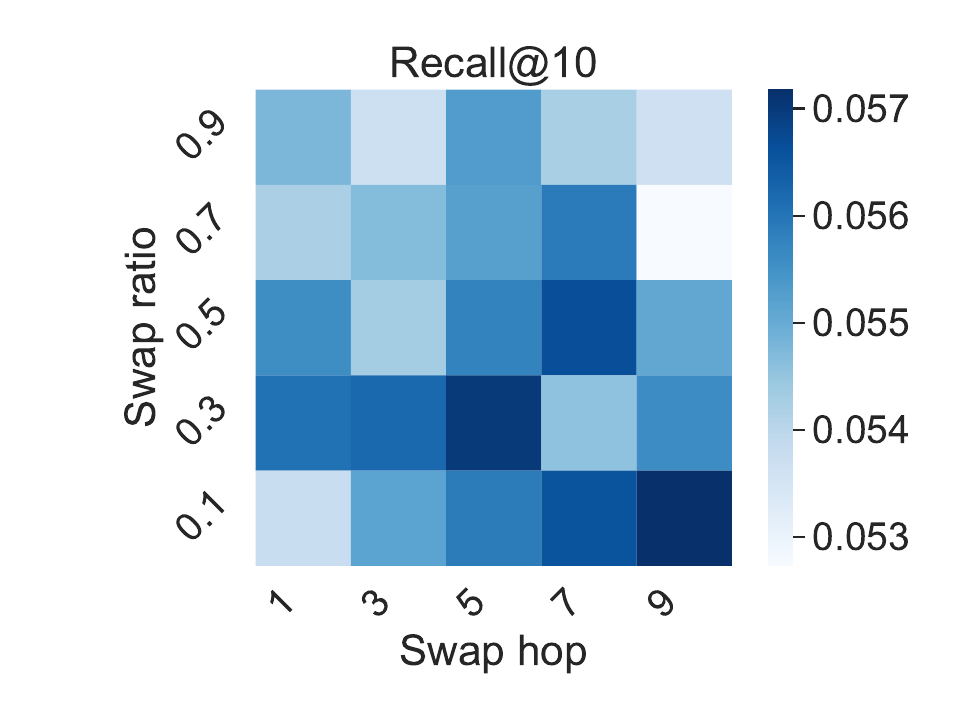}
        \vspace*{-7mm}
        \caption{{\small Instacart (item-random)}}       
    \end{subfigure}
    \hspace{-2mm}
    \begin{subfigure}[b]{0.3\linewidth}   
        \centering 
        \includegraphics[width=\linewidth]{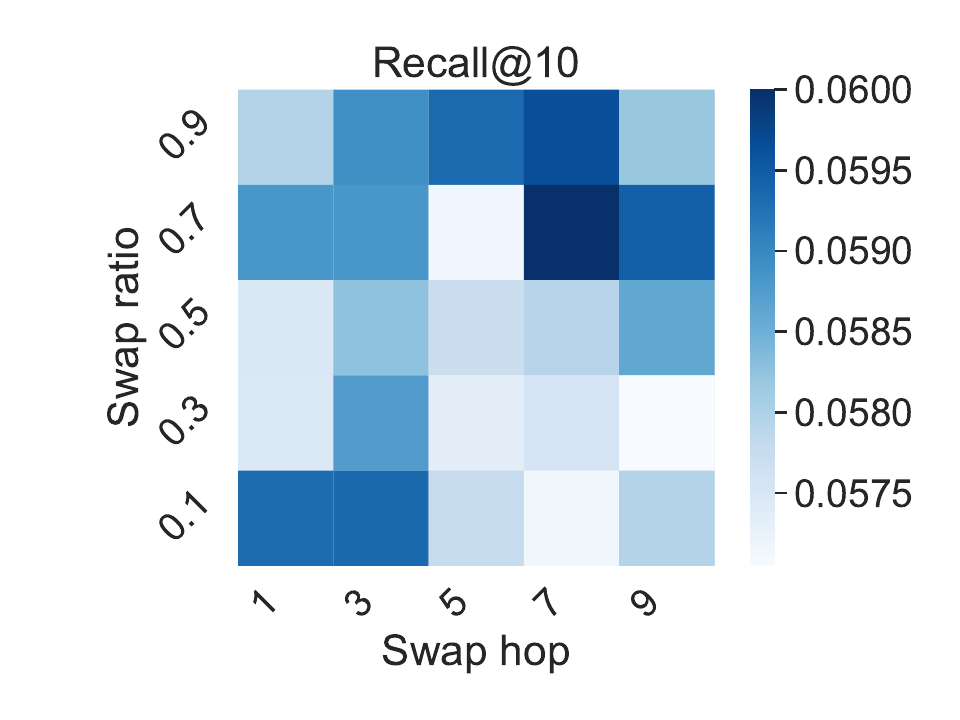}
        \vspace*{-7mm}
        \caption{{\small Instacart (item-select)}}          
        \label{fig:mean and std of net44}
    \end{subfigure}
    \caption{\small Performance heatmap with different swap hops and swap ratios.} 
    \label{fig: swap-heatmap}
\end{figure}

We further investigate the influence of hyper-parameters of the swapping strategy, i.e., swap ratio and swap hop. 
Figure~\ref{fig: swap-heatmap} shows a heatmap w.r.t. Recall@10 on different datasets when swap ratio ranges within $[0.1, 0.3, 0.5, 0.7, 0.9]$ and swap hop ranges within $[1, 3, 5, 7, 9]$.
We observe that training with both high swap ratio and swap hop (the upper-right of the heatmap) leads to poor performance on the Tafeng and Dunnhumby dataset.
When it comes to the Instacart dataset, better performance is achieved via using a high swap-hop.
The repeat ratio on Instacart dataset is high, which means that the user's interest is relatively stable and swapping across adjacent baskets will not help, so a higher swap hop is preferred to enrich item interactions within the basket on this dataset.

Given the above findings, there is a trade-off between enriching the item interactions within baskets and respecting the original temporal order information, so it is reasonable to search for the optimal swap hyper-parameters to get the highest performance on different datasets in practice.

\subsection{Effect of mask ratio and training dynamics (RQ4)}
We investigate the effect of mask ratio and analyze how the performance changes as training goes on to further understand the properties of different masking strategies.

\begin{figure*}[h]
  \centering
  \includegraphics[width=0.95\textwidth]{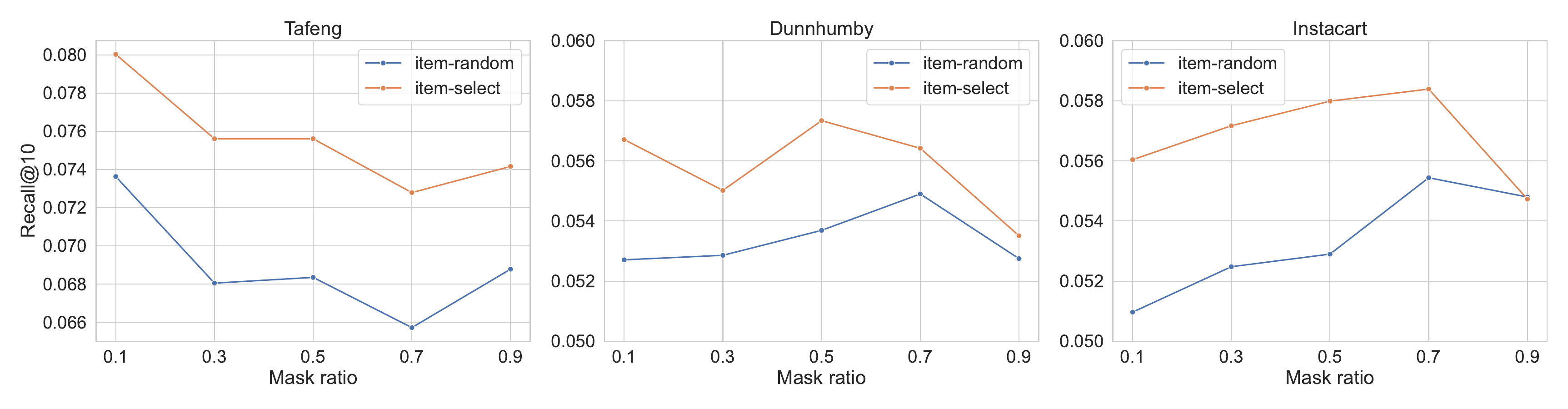}
  \vspace*{-1.5mm}
  \caption{Performance of BTBR with item-random strategy and item-select masking strategy with various mask ratios.}
  \label{fig:mask_ratio}
\end{figure*}

\begin{figure*}[h]
  \centering
  \includegraphics[width=0.95\textwidth]{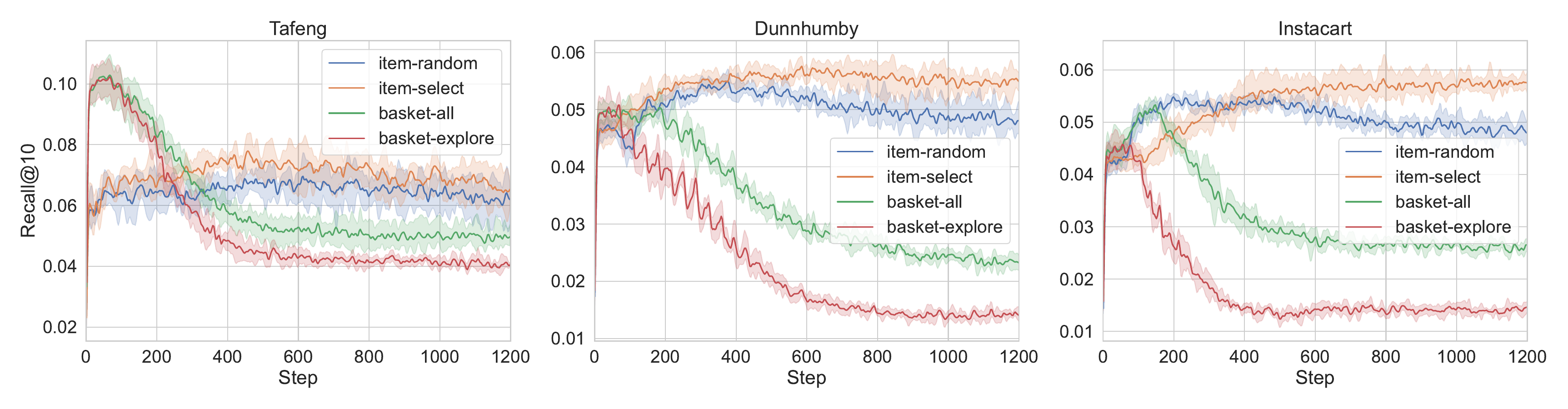}
  \vspace*{-1.5mm}
  \caption{The training progress w.r.t. Recall@10 of BTBR with different masking strategies on three datasets.}
  \label{fig:train_dynamics}
\end{figure*}

\header{Mask ratio}
The mask ratio $\alpha$ when using item-level masking is a hyper-parameter that is worth discussing. 
Figure~\ref{fig:mask_ratio} shows the Recall@10 when the mask ratio ranges within $[0.1, 0.3, 0.5, 0.7, 0.9]$. We can observe that item-select outperforms item-random with the same mask ratio in most cases. 
We also see that the optimal mask ratio is 0.1 for item-random and item-select, and the optimal mask ratio is much higher (0.5, 0.7) on the Dunnhumby and Instacart datasets. 
We suspect that a higher mask ratio is preferred in the \acs{NNBR} task when the dataset has long interaction records for the users. 



\header{Training dynamics} Figure~\ref{fig:train_dynamics} shows how the Recall@10 evolves as training goes when using different masking strategies.
First, it is obvious that basket-level masking achieves its best performances very fast, and then drops much earlier than item-level masking. This is because the training labels of basket-level masking are static, which can easily lead to overfitting, while the training labels of item-level masking are dynamic, which alleviates overfitting. 
Second, compared to basket-all masking, basket-explore masking further aggravates the overfitting problem via removing the repeat items (labels), which might lead to a performance decrease, especially in the scenario with a high repeat ratio.
Finally, the performance of item-random and item-select evolves similarly on the Tafeng dataset, since the repeat ratio on it is small. On the Dunnhumby and Instacart datasets, item-random masking results in overfitting  earlier than the item-select masking, since the masked item might still exist in other positions of the masked sequence and the model will rely more on the repeat item prediction instead of inferring novel items, as the repetition prediction task is relatively easier~\citep{nbr-rep-expl}.

\subsection{Effectiveness of joint masking (RQ5)}
So far, we have built a comprehensive understanding of different masking strategies and realize that no single masking strategy is optimal in all cases, due to the diverse characteristics of datasets.
Now, we conduct experiments to evaluate the effectiveness of joint masking (training), i.e., pre-train the model using item-select masking, then fine-tune the model using basket-all masking. The results are also shown in Table~\ref{tab:nnbr_results}.
We find that BTBR with joint masking consistently outperforms the best performance obtained by existing baselines across datasets; the improvements range from 1.3\% to 7.6\% on Tafeng dataset, from 9.2\% to 12.5\% on Dunnhumby dataset and from 19.5\% to 22.4\% on Instacart dataset. 
Joint masking does not lead to further improvements compared with a single optimal strategy, i.e., basket-all on the Tafeng dataset and item-select with swap on the Dunnhumby and Instacart datasets, in most cases.\footnote{The highest and second-highest scores in Table~\ref{tab:nnbr_results} are essentially at the same level and there is no significant difference between the joint training strategy and the single optimal strategy on each dataset in terms of performance.} The joint masking strategy under the pretrain-then-finetune paradigm is still valuable due to its robustness w.r.t. \acs{NNBR} task (i.e., it consistently achieves the best performance) on various datasets with different characteristics.


\section{Conclusion}

\iftrue
We have formulated the \acl{NNBR} task, i.e., the task of recommending novel items to users given historical interactions.
The task has practical applications, and helps us to evaluate an \acs{NBR} model's ability to find novel items for a given user. 
To understand the performance of existing NBR methods on the NNBR task, we have evaluated several \acs{NBR} models with two  training methods, i.e., Train-all and Train-explore.
To address the \acs{NNBR} task, we have proposed a bi-directional transformer basket recommendation model (BTBR), which uses a bi-directional transformer to directly model item-to-item correlations across different baskets.
To train BTBR, we have designed five types of masking strategies and training objectives considering different levels: 
\begin{enumerate*}[label=(\roman*)]
    \item item-level random masking,
    \item item-level select masking,
    \item basket-level all masking,
    \item basket-level explore masking, and
    \item joint masking. 
\end{enumerate*} 
To further improve the BTBR performance, we also  proposed an item swapping strategy to enriching item interactions. 

We have conducted extensive experiments on three datasets. 
Concerning existing \ac{NBR} methods we have found that:
\begin{enumerate*}[label=(\roman*)]
    \item the performance on the \acs{NNBR} task differs widely between existing \acs{NBR} methods;
    \item the performance of existing methods on the \acs{NNBR} task leaves considerable room for improvement, and the top performing methods on the \acs{NNBR} task are different from the top performers on the \acs{NBR} task; and
    \item training specifically for the \acs{NNBR} task by removing repeat items from the ground truth labels does not lead to consistent improvements in performance.
\end{enumerate*}
Concerning our newly proposed BTBR method, we have found that:
\begin{enumerate*}[label=(\roman*)]   
    \item BTBR with a properly selected masking and swapping strategy can substantially improve the \acs{NNBR} performance;
    \item there is no consistent best masking level for BTBR across all datasets;
    \item the proposed item-select masking strategy outperforms the conventional item-random masking strategy on the \acs{NNBR} task;
    \item the item-basket swapping strategy can further improve \acs{NNBR} performance; and
    \item a joint masking strategy is robust on various datasets but does not lead to further improvements compared to a single level masking strategy.
\end{enumerate*} 

A broader implication of our work is that blindly training specifically for the proposed recommendation task might lead to sub-optimal performance and it is necessary to consider various training objectives on diverse datasets. 
Another implication is that it is important to consider the differences between repetition behavior and exploration behavior when designing recommendation models for the grocery shopping scenario.

One limitation of this paper is that we only focus on the grocery shopping scenario. 
An obvious avenue for future work, therefore, is to extend the proposed item-select masking strategy to sequential item recommendation scenarios, and investigate if it can outperform the widely used item-random masking strategy w.r.t.\ finding novel items.
\fi



\section*{Reproducibility}

We share both our processed dataset and the source code used to produce the results in this paper at \url{https://github.com/liming-7/Mask-Swap-NNBR}.

\begin{acks}
This research was (partially) funded by the China Scholarship Council (grant \#20190607154), the Hybrid Intelligence Center, a 10-year program funded by the Dutch Ministry of Education, Culture and Science through the Netherlands Organisation for Scientific Research (NWO), \url{https://hybrid-intelligence-centre.nl}, and project LESSEN with project number NWA.1389.20.183 of the research program NWA ORC 2020/21, which is (partly) financed by the Dutch Research Council (NWO).
All content represents the opinion of the authors, which is not necessarily shared or endorsed by their respective employers and/or sponsors.
\end{acks}

\bibliographystyle{ACM-Reference-Format}
\bibliography{references}

\end{document}